\documentclass[
  twocolumn,
  pre,
  twoside,
  byrevtex,
  superscriptaddress,
  floatfix,
  longbibliography,
]{revtex4-2}

\usepackage{setspace}

\usepackage[realmainfile]{currfile}

\DeclareRobustCommand{\numngramsflipbook}{356\unskip}

\usepackage{xfrac}

\PassOptionsToPackage{hyphens}{url}\usepackage{hyperref}
\makeatletter
\g@addto@macro{\UrlBreaks}{\UrlOrds}
\makeatother

\hypersetup{
  colorlinks=true,
  allcolors=todoblue,
  urlcolor=todoblue,
  citecolor=todoblue,
  pdfborder={0 0 0},
  breaklinks=true,
}

\usepackage{array}

\usepackage{colortbl}

\makeatletter

\def\CT@@do@color{%
  \global\let\CT@do@color\relax
  \@tempdima\wd\z@
  \advance\@tempdima\@tempdimb
  \advance\@tempdima\@tempdimc
  \advance\@tempdimb\tabcolsep
  \advance\@tempdimc\tabcolsep
  \advance\@tempdima2\tabcolsep
  \kern-\@tempdimb
  \leaders\vrule
  \hskip\@tempdima\@plus  1fill
  \kern-\@tempdimc
  \hskip-\wd\z@ \@plus -1fill }
\makeatother

\usepackage[table]{xcolor}

\definecolor{olivegreen}{rgb}{0.33333,.41961,0.18431}
\definecolor{forestgreen}{rgb}{0.13333,.5451,0.13333}
\definecolor{lightgrey}{rgb}{0.7,0.7,0.7}
\definecolor{verylightgrey}{rgb}{0.90,0.90,0.90}
\definecolor{veryverylightgrey}{rgb}{0.95,0.95,0.95}
\definecolor{grey}{rgb}{0.5,0.5,0.5}
\definecolor{darkgrey}{rgb}{0.3,0.3,0.3}

\definecolor{headerblue}{HTML}{33367E}
\definecolor{unitednationsblue}{HTML}{4D88FF}

\definecolor{charcoal}{HTML}{36454F}
\definecolor{cinerous}{HTML}{98817B}
\definecolor{feldgrau}{HTML}{4D5D53}
\definecolor{glaucous}{HTML}{6082B6}
\definecolor{arsenic}{HTML}{3B444B}
\definecolor{xanadu}{HTML}{738678}

\definecolor{firebrick}{HTML}{B22222}
\definecolor{orangered}{HTML}{FF4500}
\definecolor{tomato}{HTML}{FF6347}

\definecolor{purpletaupe}{HTML}{3B444B}

\definecolor{todoblue}{RGB}{0, 91, 187}

\usepackage{graphicx,epsfig,verbatim,enumerate}
\usepackage{amssymb,amsmath}
\usepackage{ifthen}

\usepackage{longtable}

\usepackage{mathtools}

\newboolean{twocolswitch}

\newcommand{\sindex}[1]{}
\newcommand{\nindex}[1]{}

\newcommand{\www}[1]{\url{#1}}

\usepackage{lettrine}

\newcommand{\elementsymbol}{\tau}

\newcommand{\controlsymbol}{C}

\setboolean{twocolswitch}{true}

\newcommand{\onlineappendices}{Online Appendices (\href{http://compstorylab.org/trumpstoryturbulence/}{compstorylab.org/trumpstoryturbulence/})}
\newcommand{\onlineappendicesplain}{Online Appendices}
\newcommand{\arxivonly}[1]{#1}

\begin{document}

\title{\protect
Computational timeline reconstruction of the stories surrounding Trump:\\
Story turbulence, narrative control, and collective chronopathy
}

\author{
  \firstname{Peter Sheridan}
  \surname{Dodds}
}

\email{peter.dodds@uvm.edu}

\affiliation{
  Computational Story Lab,
  Vermont Complex Systems Center,
  MassMutual Center of Excellence for Complex Systems and Data Science,
  Vermont Advanced Computing Core,
  University of Vermont,
  Burlington, VT 05401.
  }

\affiliation{
  Department of Mathematics \& Statistics,
  University of Vermont,
  Burlington, VT 05401.
  }

\author{
  \firstname{Joshua R.}
  \surname{Minot}
}

\affiliation{
  Computational Story Lab,
  Vermont Complex Systems Center,
  MassMutual Center of Excellence for Complex Systems and Data Science,
  Vermont Advanced Computing Core,
  University of Vermont,
  Burlington, VT 05401.
  }

\author{
  \firstname{Michael V.}
  \surname{Arnold}
}

\affiliation{
  Computational Story Lab,
  Vermont Complex Systems Center,
  MassMutual Center of Excellence for Complex Systems and Data Science,
  Vermont Advanced Computing Core,
  University of Vermont,
  Burlington, VT 05401.
  }

\author{
  \firstname{Thayer}
  \surname{Alshaabi}
}

\affiliation{
  Computational Story Lab,
  Vermont Complex Systems Center,
  MassMutual Center of Excellence for Complex Systems and Data Science,
  Vermont Advanced Computing Core,
  University of Vermont,
  Burlington, VT 05401.
  }

\author{
  \firstname{Jane Lydia}
  \surname{Adams}
}

\affiliation{
  Computational Story Lab,
  Vermont Complex Systems Center,
  MassMutual Center of Excellence for Complex Systems and Data Science,
  Vermont Advanced Computing Core,
  University of Vermont,
  Burlington, VT 05401.
  }

\author{
  \firstname{Andrew J.}
  \surname{Reagan}
}
\affiliation{
  MassMutual Data Science, 
  Boston, MA 02110
  }

\author{
  \firstname{Christopher M.}
  \surname{Danforth}
}

\affiliation{
  Computational Story Lab,
  Vermont Complex Systems Center,
  MassMutual Center of Excellence for Complex Systems and Data Science,
  Vermont Advanced Computing Core,
  University of Vermont,
  Burlington, VT 05401.
  }

\affiliation{
  Department of Mathematics \& Statistics,
  University of Vermont,
  Burlington, VT 05401.
  }

\date{\today}

\begin{abstract}
  \protect
  Measuring the specific kind, temporal ordering, diversity, and turnover rate
of stories surrounding any given subject
is essential to 
developing a complete reckoning
of that subject's historical impact.
Here, we use Twitter as a distributed news and opinion aggregation source
to identify and track the dynamics of
the dominant day-scale stories around Donald Trump,
the 45th President of the United States.
Working with a data set comprising around 20 billion 1-grams,
we first compare each day's 1-gram and 2-gram usage frequencies
to those of a year before,
to create day- and week-scale timelines for Trump stories
for 2016--2021.
We measure Trump's narrative control, the extent to which stories
have been about Trump or put forward by Trump.
We then quantify story turbulence and
collective chronopathy---the rate at which
a population's stories for a subject seem to change over time.
We show that 2017 was the most turbulent overall year for Trump.
In 2020, story generation slowed dramatically during
the first two major waves of the COVID-19 pandemic,
with rapid turnover returning first with the
Black Lives Matter protests following George Floyd's murder
and
then later by events leading up to and following the
2020 US presidential election, including the storming of the US Capitol
six days into 2021.
Trump story turnover for 2 months during the COVID-19 pandemic
was on par with that of 3 days in September 2017.
Our methods may be applied to any well-discussed phenomenon,
and have potential to enable
the computational aspects of journalism, history, and biography.

\end{abstract}

\maketitle

\section{Introduction}
\label{sec:storyturbtrump.introduction}

What happened in the world last week?
What about a year ago?
As individuals, it can be difficult for us to freely
recall and order in time---let alone make sense of---events that have occurred
at scopes running from personal and day-to-day
to global and
historic~\cite{gunter1987a,
  price1993a,
  neisser1994a,
  lang2000a,
  tulving2002episodic,
  schacter2008searching,
  fivush2011a,
  ocasio2016a,
  garibaldi2017a,
  loftus2018a}.   One emblematic challenge for remembering story timelines is presented 
by the 45th US president Donald J.\ Trump,
our interest here.
Stories revolving around Trump have been abundant and diverse in nature.
Consider, for example,
being able to remember and then order stories involving:
North Korea,
Charlottesville,
kneeling in the National Football League,
Confederate statues,
family separation,
Stormy Daniels,
Space Force,
and
the possible purchase of Greenland.

Added to these problems of memory is that
people's perception of the passing of time is subjective and
complicated~\cite{tulving2002a,
  edy2006a,
  droit-volet2007a,
  sackett2010a,
  phillips2010a,
  grondin2010a,
  allman2012a,
  rudd2012a}. Days can seem like months (``this week dragged on forever'')
or might seem to be over in a flash (``time flies'').
Story-wise,
periods of time can also range
from being narratively simple
(``it was the only story in town'')
to complicated and hard to retell
(``everything happened all at once'').
At the population scale, major news stories may
similarly arrive at slow and fast paces,
and may be coherent or disconnected.
As one example, within the space of around 15 minutes
after 9 pm US Eastern Standard Time on March 11, 2020,
Tom Hanks and Rita Wilson announced that they had tested positive for COVID-19,
the National Basketball Association put its season on hold indefinitely due to
the COVID-19 pandemic,
and
Trump gave an Oval Office Address
during which the Dow Jones Industrial Average futures dropped.
And to help illustrate the potential disconnection of co-occurring stories
within the realm of US politics,
at the same time as the above events were unfolding,
former US vice presidential candidate Sarah Palin
was appearing on the popular Fox TV show ``The Masked Singer''
performing Sir Mix-A-Lot's ``Baby Got Back'' in a bear costume.

Here, in order to quantify story turbulence around Trump---and
the collective experience of story turbulence around Trump---we
develop a data-driven, computational approach
to constructing a timeline of stories surrounding
any given subject, with high resolution in both time
and narrative (see Data and Methods, Sec.~\ref{sec:storyturbtrump.methods}).

For data, we use Twitter as a vast, noisy,
and distributed news and opinion aggregation
service~\cite{dodds2011e,gallagher2019a,jackson2020a,tangherlini2020a,allen2020a}.
Beyond the centrality of Twitter to
Trump's communications~\cite{lee2016a-nytimes-trump-insults,ott2017a,bovet2018a,ott2019twitter,dodds2019a,ouyang2020a},
a key benefit of using Twitter
as ``text as data''~\cite{monroe2008a,grimmer2013a,gentzkow2019a,riffe2019a,boyd2020a}
is that popularity of story is encoded and recorded
through social amplification by retweets~\cite{alshaabi2020a}.
We show that Twitter is an effective source for our treatment though
our methods may be applied broadly to any
temporally ordered, text-rich data sources.

We define, create, and explore week-scale timelines of the
most `narratively dominant' 1-grams and 2-grams
in tweets containing the word Trump
(Sec.~\ref{subsec:storyturbtrump.timeline}).
We supply day-scale timelines as part of
the paper's \onlineappendices.

Having a computationally determined timeline of $n$-grams
then allows us to operationalize and measure a range of features of story dynamics.

First, and in a way particular to Trump, we quantify narrative control:
The extent to which $n$-grams being used in Trump-matching tweets are due
to retweets of Trump's own tweets (Sec.~\ref{subsec:storyturbtrump.narrativecontrol}).
We show Trump's narrative control varies from effectively zero
(e.g., `coronavirus') to high (`Crooked Hillary').

Second, we compute, plot, and investigate the normalized usage frequency time series for
$n$-grams that are narratively dominant in Trump-matching tweets for three or more days
(Sec.~\ref{subsec:storyturbtrump.timeseries}).
Along with their temporal ordering,
these day-scale time series provide a rich representation of the shapes of
Trump-related stories including shocks, decays, and resurgence.
We also incorporate our narrative control measure into 
these time series visualizations.

Third, we measure story turbulence at the scale of months by comparing
Zipf distributions~\cite{zipf1949a} for 1-gram usage frequencies between individual days
across logarithmically increasing time scales
(Sec.~\ref{subsec:storyturbtrump.storyturb}).
We are able to quantify and show, for example, that story turbulence for Trump was highest in
the second half of 2017 and lowest during the lockdown
period of the COVID-19 pandemic in the US.

Finally, we are then able to realize a numeric measurement of
`collective chronopathy' which we define
to be how time seems to be passing (`time-feel') by the rate of story
turnover (Sec.~\ref{subsec:storyturbtrump.chronopathy}).
We quantify how the past can seem more distant or closer to the present
as we move through time, and that it may do so nonlinearly
as a function of how far back we look in time.
The course of the COVID-19 pandemic in the US rendered, for example,
July 2020 as being closer to April 2020 than June.

\section{Data and Methods}
\label{sec:storyturbtrump.methods}

We draw on a collection of around 10\%
of all tweets starting in 2008.
We take all English language tweets~\cite{alshaabi2020a,alshaabi2020d}
matching the word `Trump' from 2015/01/01 on.
We ignore case and accept matches of `Trump' at any location of a tweet
(e.g., `@RealDonaldTrump' matches).
We break these
Trump-matching tweets into 1-grams and 2-grams, and create Zipf
distributions at the day scale per Coordinated Universal Time (UTC).
In previous work, we have assessed the popularity of major US political figures
on Twitter, finding that the median usage rank of the word `trump' across all of Twitter
is less than 200, tantamount to that of basic English function words (e.g., `say')~\cite{dodds2019a}.
Consequently, our resulting data set is considerable
containing around 20 billion 1-grams.

Our main collection of Trump-matching tweets thereby
includes tweets about Trump and tweets by
Trump.
Retweets and quote retweets
are naturally accounted for as they are individually
recorded in our database.
We further filter $n$-grams for simple latin
character words including hashtags and handles.

To quantify the degree to which Trump might be in control of a story,
we make a second database using the subset of $n$-grams found
in retweets of Trump's tweets (we exclude any quote tweet matter).
We then generate day-scale Zipf distributions again,
in the same format as for all Trump-matching tweets.

We perform two main analyses of these time series of Zipf
distributions,
treating 1-gram and 2-gram distributions separately.
First, we determine which $n$-grams are most
`narratively dominant' by
comparing  each day's Zipf distribution with the
Zipf distribution of a year before using our allotaxonometric instrument of
rank-turbulence divergence (RTD)~\cite{dodds2020a}.
In Fig.~\ref{fig:storyturbtrump.allotaxonometer9000-trump-turbulence-year100-example},
we provide and explain an example RTD-based allotaxonograph
to determine the narratively dominant 1-grams in Trump-related tweets
for the date of the Capitol insurrection, 2021/01/06.

We compare the top 10,000 $n$-grams and 
set the RTD parameter $\alpha$ at 1/4, a reasonable fit for Twitter data.
Tuning away from 1/4 gives similar overall results as does the use
of different kinds of probability-based divergences such as Jensen-Shannon divergence.
We use a rank-based divergence because of the plain-spoken interpretability
and general statistical robustness that ranks confer.

We find a year to be both a stable time gap with six months
to two years also producing similar results.
We then create day- and week-scale timelines of keyword $n$-grams,
filtering out hashtags and user handles.

Second, we use RTD to quantify the turbulence between Zipf
distributions for any pair of dates.
To systematically measure change in story
over time, we compare each date's Zipf distribution to that of an
approximately logarithmically increasing sequence of days $\delta$
before.

We use the one instrument of rank-turbulence divergence
throughout our paper for two reasons:
(1) Analytic coherence (each section builds out from the previous ones),
and
(2) Facilitation of quantification.
While we would expect sophisticated topic modeling approaches would 
also help elucidate stories, our goals are more expansive
regarding the experience of time.

\section{Analysis and Discussion}
\label{sec:storyturbtrump.results}

\subsection{Computational timeline generation for dominant Trump stories}
\label{subsec:storyturbtrump.timeline}

\begin{figure*}[tp!]
  \centering
  \includegraphics[height=0.80\textheight]{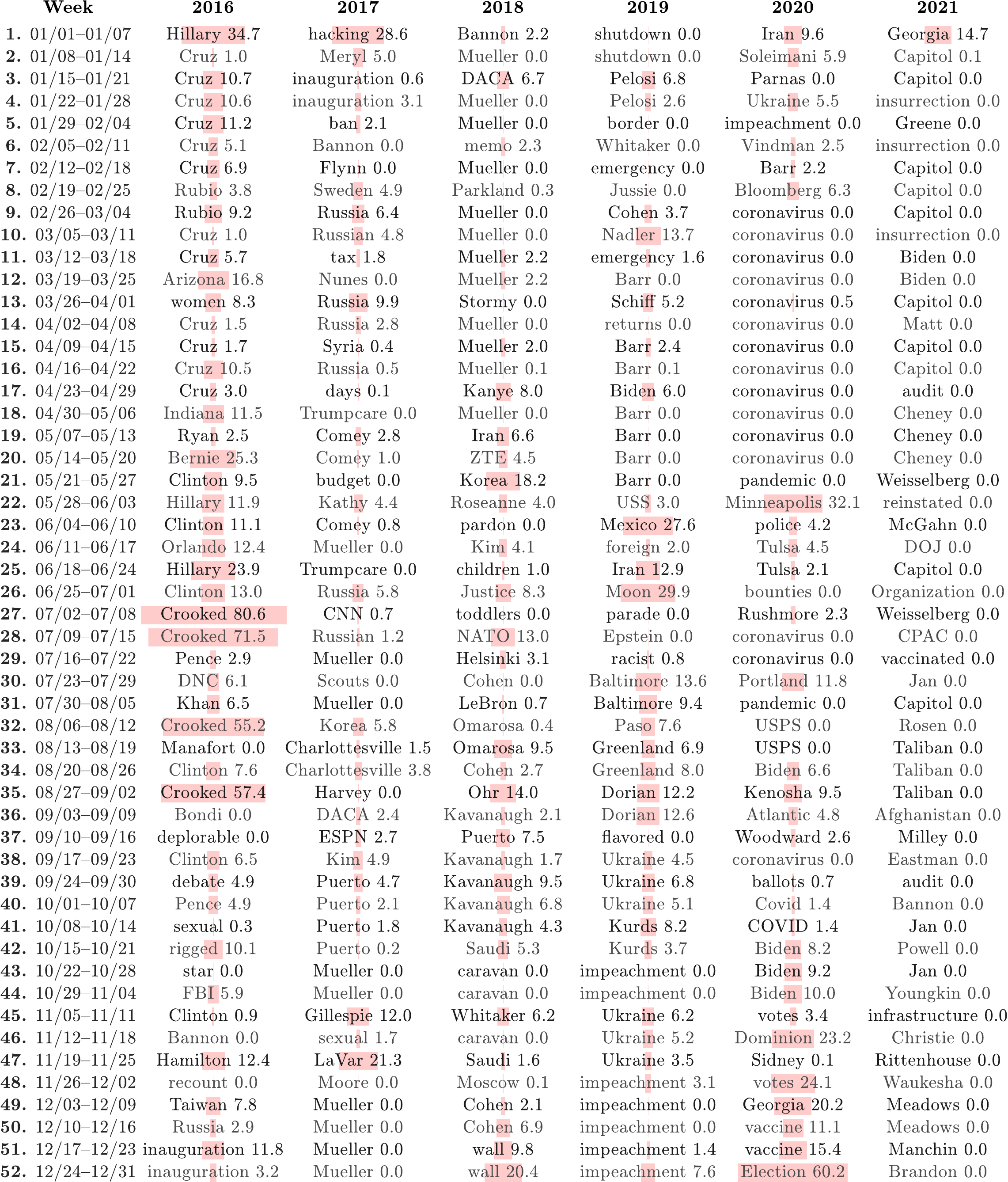}
  \caption{
    \textbf{Computational timeline of narratively dominant 1-grams surrounding Trump-matching tweets at the week scale.}
    The timelines of 1-grams and 2-grams (Fig.~\ref{fig:storyturbtrump.2grams})
    give an overall sense of story turbulence through turnover and recurrence.
    For each week, we show the 1-gram with the highest sum of
    rank-turbulence divergence (RTD) contributions~\cite{dodds2020a}
    relative to the year before,
    based on comparisons of day-scale Zipf distributions for 1-grams in Trump-matching tweets
    See Fig.~\ref{fig:storyturbtrump.allotaxonometer9000-trump-turbulence-year100-example} for
    a full example of 2021/01/06, the day of the Capitol insurrection.
    Each 1-gram must have been the most narratively dominant for at least 1 day of the week
    (i.e., highest RTD contribution).
    The numbers and pale pink bars represent Trump's narrative control as measured
    by the percentage of an $n$-gram appearing in retweets of Trump's own tweets
    during a given week.
    The limits of 0 and 100 for narrative control
    thus correspond to a 1-gram being never tweeted by Trump
    and a 1-gram only appearing in retweets of Trump.
    After Trump's account was suspended 
    by Twitter following the Capitol insurrection, 
    Trump's narrative control necessarily falls to 0.
    To align weeks across years, we assign the final 8 days to Week 52,
    and for each leap year we include February 29 as an 8th day in week 9.
    See the paper's
    \onlineappendices\
    for the analogous visualization at the day scale
    as well as a visualization of the 
    daily top 10 most narratively dominant 1-grams.
    In constructing this table
    and the table for 2-grams in
    Fig.~\ref{fig:storyturbtrump.2grams},
    we excluded hashtags,
    a small set of function word $n$-grams,
    and expected-to-be surprising $n$-grams
    such as `of the' and, in 2017, `President Trump'.
  }
\label{fig:storyturbtrump.1grams}
\end{figure*}

\begin{figure*}[tp!]
  \centering
  \includegraphics[height=0.72\textheight]{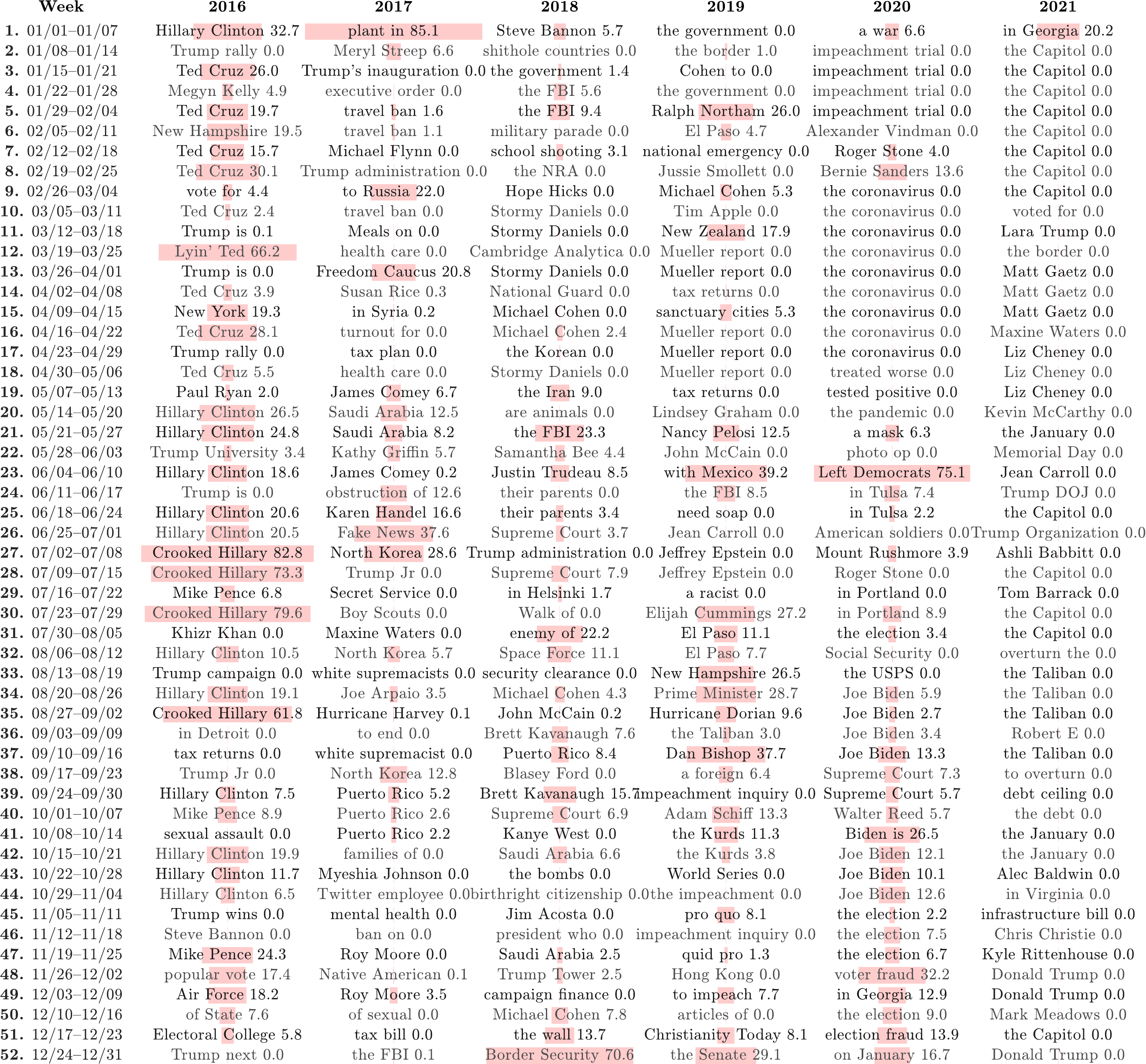}
  \caption{
    \textbf{Computational timeline of narratively dominant 2-grams surrounding Trump at the week scale.}
    The timeline's construction and the plot's details are
    analogous to that of Fig.~\ref{fig:storyturbtrump.2grams}.
    The $n$-grams in both timelines
    match (`coronavirus' and `the coronavirus'),
    expand on each other (`bounties' and `American troops'),
    or point to different stories (`Mueller' and `Stormy Daniels').
    See the paper's
    \onlineappendices\
    for a day-scale visualization of the top 10
    most narratively dominant 2-grams in Trump-matching tweets.
  }
\label{fig:storyturbtrump.2grams}
\end{figure*}

In Figs.~\ref{fig:storyturbtrump.1grams} and~\ref{fig:storyturbtrump.2grams},
we show computational
timelines of the most narratively dominant 1-grams and 2-grams
in Trump-matching tweets
for each week running from the start of 2016 into 2020.
We coarse-grain from days to weeks by finding $n$-grams with
the highest overall rank-turbulence divergence (RTD) sum for each week.

Our computational timelines provide $n$-grams as keyword hooks,
and immediately give a rich overall view of the major stories
surrounding Trump.
Broadly, the early major chapters run through the Republican nomination process,
the election, and inauguration.
Reflected in names of individuals and entities,
places, and processes,
the timelines then move through a range of
US and world events
(North Korea, Charlottesville, Parkland, Iran,
George Floyd's murder, Portland);
US policy and systems (travel ban, Space Force, southern border wall, Supreme Court);
natural disasters (hurricanes in 2017, COVID-19 in 2020);
scandals (Russia, Stormy Daniels, Mueller, impeachment,
Taliban bounties for American soldiers),
and the 2020 US election and aftermath
(death of Ruth Bader Ginsburg,
debate with Biden, Trump's contraction of COVID-19,
claims of fraud, a focus on Georgia, the storming of the US Capitol).

The week-scale timelines for 1-grams and 2-grams 
variously directly agree (e.g., `Epstein' and `Jeffrey Epstein' in week 27 of 2019,
and `coronavirus' and `the coronavirus' for 9 consecutive weeks in 2020);
make connections (e.g., `Crooked', `Hillary', and `Crooked Hillary' in 2016);
or point to different stories
(e.g., `Syria' and `Trump Foundation' in week 51 of 2018).

The $n$-gram timelines also provide an overall qualitative
sense of story turbulence.
In the lead up to the 2016 election,
the stories around Trump largely concern his opponents,
particularly Ted Cruz and Hillary Clinton.
Near the election, we see an increase in story turbulence
with `sexual assault', `rigged', and `FBI'.
During Trump's presidency, the timelines give
a sense of stories becoming more enduring over time.
After a tumultuous first nine months of 2017,
we see the first stretch of four or more weeks
being ruled by the same dominant story:
`Puerto'/`Puerto Rico' and `Mueller'.
In 2018, `Mueller' then dominates for 12 of 17 weeks
and then later `Kavanaugh' leads for 5 of 6 weeks.
In 2019, `Barr' is number one for 4 of 5 weeks early on.
Later, `Ukraine' and `impeachment' reflect the main story
of the last four months of 2019, broken up only by `Kurds'.
And in 2020, COVID-19 is the main story for 13 consecutive weeks
for 1-grams starting at the end of February,
pushed down by the murder of George Floyd and subsequent protests,
before returning again as the major story of July.

Our intention with
Figs.~\ref{fig:storyturbtrump.1grams} and~\ref{fig:storyturbtrump.2grams}
is to give one page summaries of over five years of stories around Trump.
Below these dominant $n$-grams at the week scale are
$n$-grams at the day scale representing many other stories.
For example,
Tab.~\ref{tab:storyturbtrump.troops}
shows the the top 12 most narratively dominant 1-grams for
three consecutive days in late June of 2020 where
the story of Russia paying Taliban soldiers bounties
for killing US soldiers rose abruptly to prominence.
We see that by June 27, references to the
COVID-19 pandemic had dropped down the list
(`coronavirus' would return to the top 3 on July 1),
as had indicators of
other stories (`Biden', `lobster', and `statues' point to
Trump's 2020 challenger,
a call by Trump to subsidize the US lobster industry,
and the taking down of Confederate statues
in the aftermath of George Floyd's murder).
The Russian bounties story would fall away within a week,
with the next non-pandemic story appearing being
the arrest of Ghislaine Maxwell on July 2, 2020.
As part of the \onlineappendicesplain,
we provide a range of visualizations and data files
for day-scale computational timelines
for the most narratively dominant 1-grams and 2-grams.

\begin{table}[tp!]
  \setlength{\tabcolsep}{0.5em}
  \renewcommand{\arraystretch}{1.2}
  \rowcolors{2}{white}{gray!15}
  \begin{tabular}{cccc}
    Rank &     2020/06/25 & 2020/06/26 & 2020/06/27 \\
    \hline
    1   & coronavirus &    pandemic   &    bounties    \\
    2   & Biden       &    bounties   &    bounty      \\
    3   & pandemic    &    coronavirus&    soldiers    \\
    4   & testing     &    Biden      &    militants   \\
    5   & lobster     &    militants  &    Russia      \\
    6   & Matter      &    hiring     &    kill        \\
    7   & statues     &    cases      &    briefed     \\
    8   & virus       &    testing    &    Afghanistan \\
    9   & cases       &    virus      &    intelligence\\
    10   & ad          &    Statues    &    Taliban     \\
    11   &  fishing     &    bounty     &    troops      \\
    12   & ChinaVirus  &    Care       &    pandemic    \\
  \end{tabular}
  \caption{
    Top 12 most narratively dominant 1-grams for Trump-matching tweets
    for an example three day period.
    The sharp rise of the Russian bounties for Taliban soldiers to kill US troops,
    along with the fall of other stories,
    gives an example of a microscopic
    detail of story turbulence we aim to quantify macroscopically.
  }
  \label{tab:storyturbtrump.troops}
\end{table}

\subsection{Narrative control}
\label{subsec:storyturbtrump.narrativecontrol}

Because we are working with Twitter,
some $n$-grams may become narratively dominant
due to high overall use by many distinct individual users
because of one highly retweeted individual's tweet,
and all degrees in between.
Trump is the evident potential source of many retweets in our data set,
and we introduce the concept of `narrative control' as part of our analysis.
In Figs.~\ref{fig:storyturbtrump.1grams} and~\ref{fig:storyturbtrump.2grams},
the underlying pale pink bars and accompanying numbers
indicate our measure of Trump's narrative control,
$\controlsymbol_{\elementsymbol,t}$.
We quantify
$\controlsymbol_{\elementsymbol,t}$
for any given $n$-gram $\elementsymbol$ on
Twitter for some time period $t$
as the percentage of the occurrences of
$\elementsymbol$ due to retweets of Trump tweets:
$\controlsymbol_{\elementsymbol,t}
=
100
f_{\elementsymbol,t}^{\textnormal{rt}}/
f_{\elementsymbol,t}$.

We see that Trump's week-scale
levels of narrative control vary greatly over time.
A few example highs, ordered by their date of
first becoming narratively dominant, are
`Crooked Hillary' (82.6\%),
`Fake News' (37.6\%)
`Border Security' (70.6\%),
`Minneapolis' (32.1\%),
and
`Left Democrats' (75.1\%).

By contrast Trump's narrative control has been low to non-existent
for narratively dominant $n$-grams representing many stories.
For `Mueller' in 2017 and 2018,
$\controlsymbol_{\elementsymbol,t}$
ranges from 0\% to 2.2\%
(see however Sec.~\ref{subsec:storyturbtrump.timeseries} below).
The names
`Stormy Daniels',
`Jeffrey Epstein',
and
`Ghislaine Maxwell'
all register
$\controlsymbol_{\elementsymbol,t}$=0,
as does `the bombs' in reference to pipe bombs
mailed to leading Democrats and journalists in October, 2018.

\begin{figure*}[tp!]
  \includegraphics[width=\linewidth]{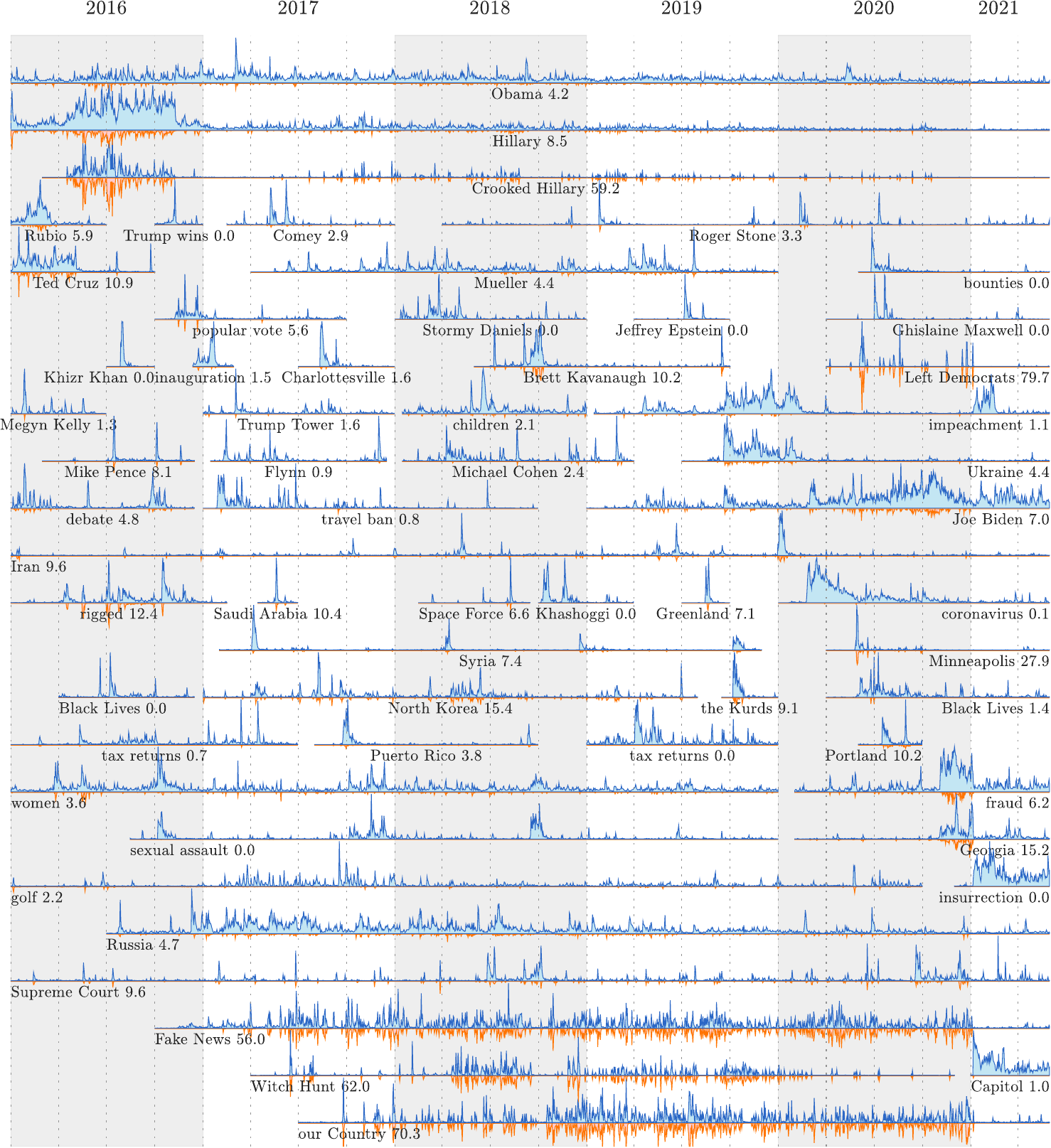}
  \caption{
    Example time series of narratively dominant 1-grams and 2-grams
    for Trump-matching tweets.
    Each time series shows the relative usage rate for $n$-grams (blue curve),
    and Trump's level of narrative control (inverted orange curve with pale pink fill).
    See the
    \onlineappendices\ \arxivonly{(or the anciliary files section for the paper's arXiv version)}
    for full time series
    for all $n$-grams which have been narratively dominant on at least three days.
    The background shading and vertical dashed lines indicate
    years and quarters.
  }
  \label{fig:storyturbtrump.major_story_timelines_trump001}
\end{figure*}

Strikingly, Trump's narrative control
for `coronavirus' has been effectively 0.
Of course, and as for all $n$-grams,
Trump may have used other terms to refer to them.
In the case of `coronavirus', he has used `corona virus'
(with optional capitalizations), `invisible enemy',
and a derogatory term.
But our measurement here gets at the degree
to which Trump has engaged with a specific $n$-gram
being used on Twitter in connection with him.

Transitions in narrative control also stand out.
In 2020, the most striking shifts at the week scale
are 0\% for `pandemic' to 32.1\% for Minneapolis,
0\% for `photo op' to 75.1\% for `Left Democrats',
and 0\% for `coronavirus' to 11.8\% for `Portland'.
And the start of 2020 with the US assassination
of Iranian general Soleimani jumps out amid
the $n$-grams of the impeachment hearings 
(`Iran' 9.6\%,
`Soleimani' 5.9\%,
`a war' 6.6\%).

The storming of the Capitol on 2021/01/06 by
Trump supporters lead to Twitter
permanently banning Trump's Twitter account.
In the context of Twitter and by the nature of our measure,
Trump's narrative control therefore
ended abruptly on 2021/01/08.

\subsection{Time series for dominant Trump stories}
\label{subsec:storyturbtrump.timeseries}

We now take our list of narratively dominant 1-grams and 2-grams
and extract day-scale
time series of normalized usage rate in Trump-matching tweets.
We also generate the corresponding narrative control time series.

In Fig.~\ref{fig:storyturbtrump.major_story_timelines_trump001},
we show a selection of $n$-gram time series (blue lines)
with Trump's narrative control time series inverted (orange lines with
pale pink fill).
Below each horizontal axis,
we annotate the $n$-gram along with the overall
narrative control percentage for section of time series visible.
We normalize each time series by its maximum.

We see a range of motifs common to sociotechnical time series:
spikes (`Saudi Arabia', `Syria', `Cambridge Analytica'),
shocks with decays (`Puerto Rico', `Kurds'),
sharp drop offs (`Ted Cruz'),
episodic bursts (`Roger Stone', `Syria', `sexual assault'),
and noise
(`Russia', `Obama')~\cite{crane2008a,dewhurst2020b,dedomenico2020a}.

While an $n$-gram may be ranked as the most narratively dominant
for some period of time, its
usage rate may fluctuate during that time.
A clear example is the time series for `coronavirus' which begins with
an initial spike followed by a shock, and then trends linearly down until the end of June, 2020.

As it is derived from a subset's fraction of a whole,
the narrative control time series can at most exactly mirror the overall time
series.
Four examples of time series with high and enduring narrative control percentages
are `Crooked Hillary', `Witch Hunt', `Fake News',
and `our Country'
(59.2\%, 62.0\%, 56.2\%, and 70.7\%).

More subtly, we see that Trump's narrative control for `Mueller'
increases over time,
though remaining modest at 4.4\%.

Many of the examples we show exhibit relatively flat narrative control time series.
For 2-grams, Trump did not tweet about `Black Lives' (Matter)
in 2016, and only a fraction in 2020 (2.5\%).
And we see zero narrative control for
`sexual assault'
whose time series correlates strongly with `women'.

Post the 2020 election,
we show three time series connected with
Trump's claims of a `rigged election':
`fraud' (6.6\%),
`Georgia' (16.3\%),
and `Capitol' (2.0\%).

In the \onlineappendicesplain\ \arxivonly{and as an anciliary file on the arXiv},
we include
a PDF booklet of time series for all $n$-grams
which have been narratively dominant in Trump tweets
on at least three days.
We extend the time series for all $n$-grams to run from 2016/01/01 on,
and order them by the date they first achieve narrative dominance.
Moving through this ordered sequence of
\numngramsflipbook\
time series gives another qualitative
experience of how the stories around Trump have unfolded in time,
placing them in temporal context.

\begin{figure*}[tp!]
  \includegraphics[width=\textwidth]{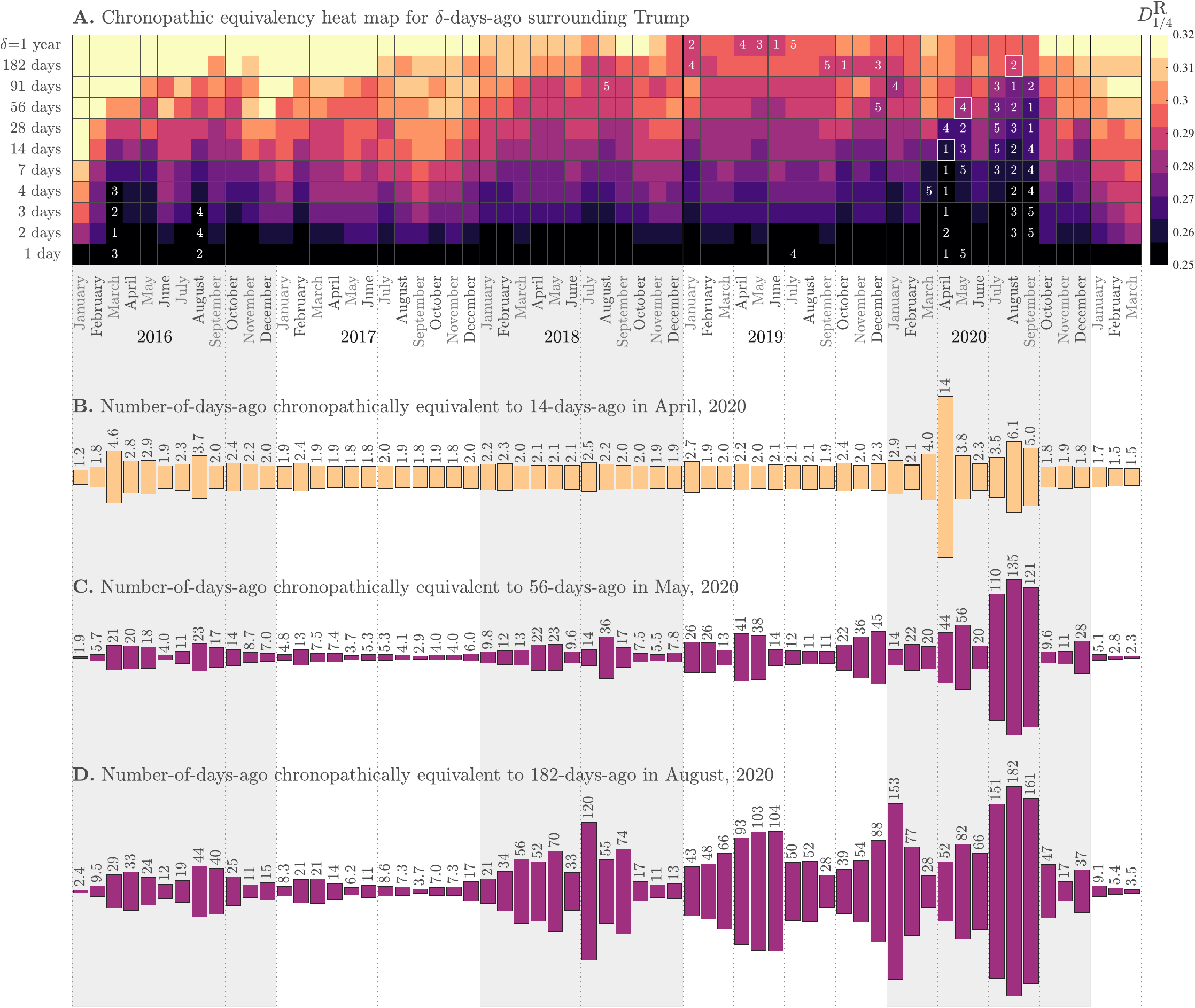}
  \caption{
    \textbf{A.}
    Chronopathic equivalency heat map: Each cell represents the `story distance'
    for a given month and a given number of $\delta$ days before.
    We measure story distance as
    the median rank-turbulence divergence (RTD) between the Zipf distributions
    of 1-grams used in Trump-matching tweets for
    each day of a month
    and $\delta$ days before (we use RTD parameter $\alpha$=1/4).
    Lighter colors
    on the perceptually uniform color map
    correspond to higher levels of story turnover.
    Numbers indicate the slowest five months for each value of $\delta$.
    After the story turbulence of the 2016 election year
    and especially the first year of Trump's presidency,
    there has been a general slowing down in story turnover at all time scales
    (the `plot thickens').
    By September US of 2020, the COVID-19 pandemic had induced record
    slowing down of story turnover around Trump at time scales up to 91 days,
    the story being punctuated by and then combined with
    the Black Lives Matter protests in response to
    George Floyd's murder on May 25, 2020.
    \textbf{B.}
    Using an example anchor of April 2020 and $\delta$=14 days
    (white square in panel \textbf{A}),
    a plot of chronopathic equivalent values of $\delta$ across time.
    During Trump's presidency, the same story turnover occurred
    as fast as every 1.8 days in September 2017 and
    1.7 days in October 2020.
    Because story turbulence is nonlinear,
    using a different anchor month and $\delta$ (i.e., selecting
    a different cell in the heatmap) potentially gives a different
    chronopathic equivalency plot.
    \textbf{C.}
    Anchor of 56 days in May 2020.
    \textbf{D.}
    Anchor of 182 days in August 2020.
    See the \onlineappendices\ for the most recent version of this figure.
    As for Fig.~\ref{fig:storyturbtrump.major_story_timelines_trump001},
    shading and lines give guides for years and quarters.
  }
  \label{fig:storyturbtrump.rank_turbulence_divergence_timeseries_trump_timewarp050}
\end{figure*}

\subsection{Story turbulence}
\label{subsec:storyturbtrump.storyturb}

We have so far demonstrated that the $n$-grams of Trump-matching tweets
track major events and narratives around Trump.
We now  move to using our comparisons of Zipf distributions
for 1-grams via rank-turbulence divergence (RTD) to operationalize
and quantify two aspects of story experience:
(1) Story turbulence, the rate of story turnover surrounding Trump,
and
(2) Collective chronopathy, the feeling of how time passes at a population scale.
Chronopathy is to be distinguished from the differently defined chronesthesia~\cite{tulving2002a}.
Here, we use the -pathy suffix primarily to mean `feel' (as in ` empathy')
though a secondary connotation of sickness proves serviceable too (as in `sociopathy').

In Fig.~\ref{fig:storyturbtrump.rank_turbulence_divergence_timeseries_trump_timewarp050},
we present visualizations of the collective
chronopathy for Trump-matching tweets.
We focus first on
the heat map of Fig.~\ref{fig:storyturbtrump.rank_turbulence_divergence_timeseries_trump_timewarp050}A,
the core distillation of our measurement 
of the passing of time.

For each day in each month, we compare 1-gram Zipf distributions
to the 1-gram Zipf distribution for $\delta$ days before using RTD with $\delta$ varying
approximately logarithmically from 1 day to 1 year
(vertical axis of Fig.~\ref{fig:storyturbtrump.rank_turbulence_divergence_timeseries_trump_timewarp050}A).
We generate the heat map using the median value of RTD for each month and each $\delta$.
We employ a 10 point, dark-to-light perceptually uniform color map for increasing RTD
which corresponds to the speeding up of story turnover.
For each $\delta$, we indicate the five slowest months by annotation with 1 being the slowest.

Overall, we see a general trend of story turnover slowing down---the plot thickens---with
several marked exceptions, leading to an extraordinary
slowing down across many time scales throughout the COVID-19 pandemic in 2020.

We describe the main features of the heat map, moving from left to right through time.
As we would expect, the election year of 2016 shows fast turnover at
the longer time scales of $\delta \sim$ six months to a year.
This turnover carries through into 2017 as the shift to being president
necessarily leads to different word usage around Trump.
For the shorter time scales,
January 2016 has fast turnover at all time scales---Trump's narrative
was rapidly changing as he was becoming
the lead contender in the Republican primaries.
Some of the slowest turnovers for 1 day out to 14 days
occur in 2016, notably in March and August.

Story turnover generally increases through 2017 at all time scales $\delta$,
with September, 2017 being the month with days least connected to all that had come
before in the previous year.
As suggested by Figs.~\ref{fig:storyturbtrump.1grams} and~\ref{fig:storyturbtrump.2grams},
the first year of Trump's presidency burned through story,
with just some of the dominant narratives concerning
Flynn, Comey, Russia, Mueller, North Korea, Charlottesville, DACA, NFL,
and three major hurricanes.

In 2018 and 2019, we see some of the slowest turnover at longer time scales.
Trump has the consistency of being president in the previous year.
Two periods where story turnover for shorter time scales slow down in the second
and third years of the Trump presidency are around the government shutdown
(January 2019) and the impeachment hearings (last three months of 2019).

In March 2020, the COVID-19 pandemic brings story turnover to a functional halt
at shorter time scales.
April is especially slow with either the slowest or second slowest turnovers
for time scales ranging from 1 day to 14 days.
The same time scales for May are all still slow (top 5) but now longer time scales
show less story turnover.
The murder of George Floyd on May 25 and subsequent Black Lives Matter
protests then leads to a sharp transition in the heat map.
At the same time, the pandemic had subsided before what would be new surge in August,
and was for much of June a secondary story.

Entering July 2020, story turnover slows dramatically for a second time---now slowing
at all time scales---as
the pandemic once again becomes a major narrative
along with that of the Black Lives Matter protests.
August has the slowest 91 day story turnover,
and September the same for 28 and 56 day time scales.
The text around Trump in August 2020 is closer to that of May 2020
than for any other three month comparison.

Story turnover rises sharply again in October and November of 2020,
in the lead up to and aftermath of the 2020 US presidential election.
The narratives around Trump begin to roil
at the end of September which saw the contentious first presidential debate
and the death of Justice Ruth Bader Ginsberg.
The subsequent nomination and confirmation
of Amy Coney Barrett for the Supreme Court, and the direct connection to
Trump's contraction of COVID-19, hospitalization, and recovery
then led to a thrashing of the narrative timeline.
Post election, the lack of a clear immediate winner and then
Trump's refusal to concede and claims of fraud
made November, December, and January
especially turbulent at longer time scales.

\subsection{Collective chronopathy}
\label{subsec:storyturbtrump.chronopathy}

We turn finally to determining what we will call
chronopathically equivalent time scales.
For any given month and time scale $\delta$---any cell in the heat map
of Fig.~\ref{fig:storyturbtrump.rank_turbulence_divergence_timeseries_trump_timewarp050}A---we can
estimate time scales in other months with corresponding values of RTD.

For three examples, the white-bordered squares in 2020 in
Fig.~\ref{fig:storyturbtrump.rank_turbulence_divergence_timeseries_trump_timewarp050}A
mark anchor time scales and months of
$\delta$=14 days (2 weeks) in April,
$\delta$=56 days ($\sim$ 2 months) in May,
and
$\delta$=182 days ($\sim$ 6 months) in August.
For these three anchors,
Figs.~\ref{fig:storyturbtrump.rank_turbulence_divergence_timeseries_trump_timewarp050}B--D,
show the equivalent values of $\delta$ across all months.

Fig.~\ref{fig:storyturbtrump.rank_turbulence_divergence_timeseries_trump_timewarp050}B
shows that two weeks in April felt like the longest two weeks across
the whole time frame.
All other months achieve the same level of story turnover in less than
a week, with a maximum of 6.1 days in August 2020.
With the first wave of the pandemic unfolding,
the stories around Trump became stuck and time dragged, collectively.

Lifting up to the 56-day anchor in May 2020,
in Fig.~\ref{fig:storyturbtrump.rank_turbulence_divergence_timeseries_trump_timewarp050}C,
we see the slowdown of July, August, and September 2020 dominate,
with time scales doubling.
We see some near equivalent time scales in 2018 and 2019,
with the impeachment's progression by the end of 2019 producing a 45 day period.
The equivalent time scales of around 9.5 for October and November of 2020
point to speed-up factors of $\sim \times$12.

Stepping further out to the level of story turnover in six months relative to August 2020,
Fig.~\ref{fig:storyturbtrump.rank_turbulence_divergence_timeseries_trump_timewarp050}D
We see episodic slowdowns in 2018 and 2019,
with two
main causes in 2019 being the Mueller report and the impeachment hearings.
With this anchor, we see November 2020 separate from October with
a chronopathically equivalent time scale of 13 days versus 47.

Taken together,
Figs.~\ref{fig:storyturbtrump.rank_turbulence_divergence_timeseries_trump_timewarp050}B--D
make clear, and quantify, the rapidity of story turnover in 2017,
especially September of that year.

Story turbulence is complicated as the speed up or slow down of story can
vary nonlinearly across different memory time scales.
Annual days of the year may seem close to
that of a year before (e.g., Thanksgiving) but further away
from time periods in between.
For Trump stories, our measure of story turnover across 56 days
looking back from May 2020
was equivalent to around 3 days in September 2017,
a speed up of $\times$19,
and 110 days in July 2020,
a slowing down by a factor of $\times$2.
In choosing 28 days instead of 56 in May 2020,
we would find equivalent time scales of 2.1 and 16 in September 2017 and July 2020,
with speed-up factors of $\times$13 and $\times$1.7.
So by this shorter time scale benchmark of about a month in May,
story seems to have moved faster in July
because of the narrative dominance of protests in June.

\section{Concluding remarks}
\label{sec:storyturbtrump.concludingremarks}

We have shown that interpretable, high-level timeline summaries of
historical events can be derived from Twitter.
Our process, which we have worked to make as rigorous as possible,
is necessarily computational: In our approximately 10\% sample of all of Twitter,
there were from 2015/01/01 through to 2020/08/12
around 20 billion distinct 1-grams in Trump-matching tweets,
with 1.5 billion of those being contained in retweets of Trump's tweets.

Because we have shown that the narratively dominant $n$-grams our method find are
historically sensible, we are then able to defensibly quantify
narrative control, story turnover, and collective chronopathy
in the context of Trump.

We observe that our focus on $n$-grams has resulted in timelines
that are largely descriptive of major events and stories
through the surfacing of the names of people, places, institutions, processes,
and social phenomena.
While the collective construction of what matters cannot be said to be
objective---what matters socially is what matters socially~\cite{berger1991a,sandu2016social}---we find
that our computationally derived
timelines are not laden with overt opinion or framing with the notable
exception of certain phrases due to Trump himself.

Our study of stories around Trump suggests much future possible research.

Broader explorations of Trump's (and others') narrative control should be possible.
We have limited ourselves here to narratively dominant $n$-grams
and only within the same time frame.
We have elsewhere examined ``ratioing'' of tweets by Obama and Trump~\cite{minot2021b}, 
exploring in particular the balance of likes to retweets, 
and these and other measures of interactions could provide a more
nuanced way to operationalize narrative control.
Studies across different news outlets and social media platforms
could explore the extent to which Trump's narratives are driven by external stories
and vice versa, as well as their persistence in time,
and certain modeling and prediction may be possible.

We have indexed stories by keywords of 1-grams and 2-grams.
Connecting these $n$-grams to Wikipedia would possibly allow
for the automatic generation of timelines augmented
by links to (or summaries of) Wikipedia entries.
Disconnects between Twitter and Wikipedia (and other texts) would
be of interest to explore as well.

Computationally generating a taxonomy of story type would be a natural way to
improve upon our work.
How have the kinds of stories around Trump changed over time,
which ones have persisted,
and which ones have been more likely to last only a few days?

We have presented narratively dominant $n$-grams
for the time scales of weeks, while reserving the top 20
at the day scale for the paper's \onlineappendicesplain.
While this is reasonable for a study going across now five years,
for certain major events, shorter time scales of hours or even minutes
may be better suited as the temporal units of analysis.

The relationship between Trump's favorability polls
with story turbulence and story kind could also be examined.
As a rough observation, we note that the speeding up and then slowing down of story turnover
around Trump through 2017 and into 2018 appears to be anti-correlated
with Trump's approval ratings during that same period.
Having established a story taxonomy would be important for moving in this direction.

While our specific instrumental focus has involved
the use of Twitter to analyze stories surrounding Trump,
in doing so we have laid out a general, structured approach to
quantifying and exploring story turbulence 
for any well-defined topic in temporally ordered, large-scale text
(e.g., news outlets, online forums, and Reddit).
Wherever Zipf distributions can be derived,
our methods will allow for the computational construction
of $n$-gram-anchored story timelines along with
measures of story distance, turbulence, and chronopathy.
For such extensions, we caution that the popularity of text
must be incorporated in some fashion
through measures of readership, shares, etc.
We also observe that stronger variations in findings
would potentially come from
choosing different text sources rather than
through reasonable adjustments of
the methods of analysis we have laid out here.

We believe that our techniques will be of value
for computationally augmented journalism and the computational social
sciences~\cite{lazer2009a,conte2012a,flew2012a,coddington2015a}.
News reporters and historians in particular
are faced with having to process ever more data, text, and media,
and we see our work here as contributing to a far broader effort 
to develop ways to provide powerful computational support for
the vital work performed by domain-knowledge experts.

\medskip

\acknowledgments
The authors are grateful for the computing resources provided by the
Vermont Advanced Computing Core
which was supported in part by NSF award No. OAC-1827314,
financial support from the
Massachusetts Mutual Life Insurance Company and Google Open Source
under the Open-Source Complex Ecosystems And Networks (OCEAN) project.

\clearpage

\onecolumngrid
\appendix

\renewcommand{\thefigure}{A\arabic{figure}}
\renewcommand{\thetable}{A\arabic{table}}
\setcounter{figure}{0}
\setcounter{table}{0}

\begin{figure*}[tp!]
  \includegraphics[width=\textwidth]{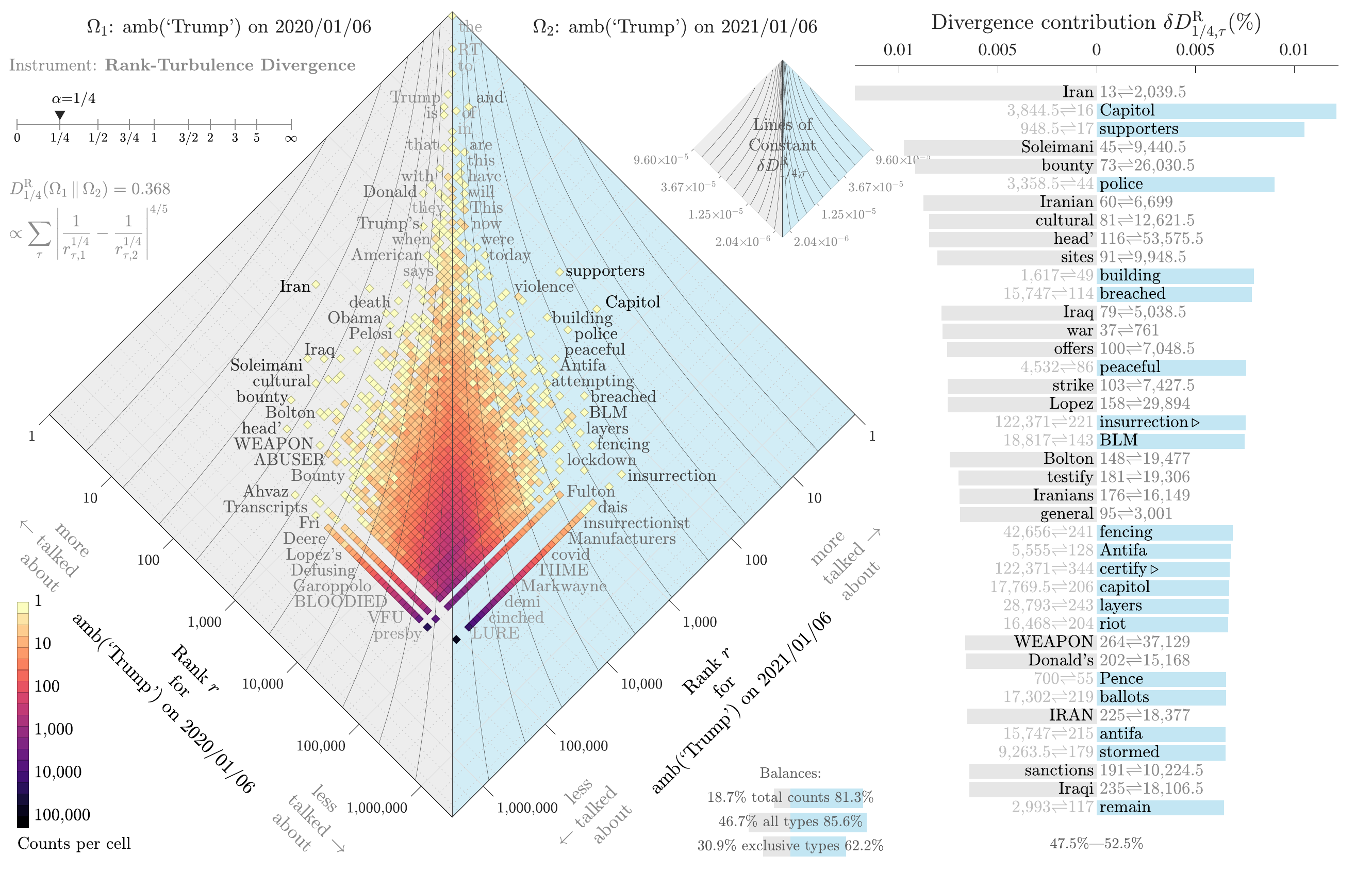}
  \caption{
    Example allotaxonograph~\cite{dodds2020a} comparing 1-grams in Trump-matching tweets
    on two dates, separated by a year: 2020/01/06 and 2021/01/06.
    Our findings all build out from allotaxonometry:
    The principled measurement of the difference between
    the architectures of any two complex systems
    as represented by Zipf distributions~\cite{zipf1949a}.
    For 2020/01/06 versus 2021/01/06, 1-grams on the first date reflect the aftermath of
    the assassination of Iranian military commander Qasem Soleimani by the United States on 2020/01/03,
    while 1-grams appearing on the second date revolve around
    the Capitol insurrection by Trump supporters.
    The rank-rank histogram on the left of the allotaxonograph displays the joint Zipf distribution,
    naturally laid out in double-logarithmic space.
    We generate the 1-gram ranking on the right of the allotaxonograph
    using rank-turbulence divergence (RTD)~\cite{dodds2020a}
    with tuning parameter
    $\alpha$=1/4.
    The contour lines on the histogram provide guides for RTD showing a reasonable fit
    to the joint Zipf distribution's form.
    Variations around $\alpha$=1/4 will not change the overall findings (i.e., the orderings of
    contributing 1-grams as well as the overall RTD score).
    See Ref.~\cite{dodds2020a} for a full explanation of allotaxonometry and RTD.
    In Sec.~\ref{subsec:storyturbtrump.timeline} in the main paper,
    we use RTD at the year scale to determine
    narratively dominant 1-grams and 2-grams arising on the second of the two dates being compared.
    By comparing to a year ago, we are able to generate a background Zipf distribution
    that will help remove calendrical features as well as generically
    Twitter- and Trump-related 1-grams (e.g., `RT' and `Donald').
    While the allotaxonograph shows 1-grams on both dates,
    we emphasize that our focus is on the second date, in that we are seeking
    to determine the most important 1-grams of today.
    For 2020/01/06 versus 2021/01/06, looking at the ranked list on the right,
    we see that the top five 1-grams for the day of the Capitol insurrection are
    `Capitol',
    `supporters',
    `police',
    `building',
    and
    `breached'.
    The salience of `BLM' (Black Lives Matter) and `Antifa'
    point to the immediate
    confusion and disinformation surrounding the Capitol insurrection.
    We provide all year-scale allotaxonographs
    at the paper's \onlineappendicesplain.
    In Secs.~\ref{subsec:storyturbtrump.storyturb} and
    \ref{subsec:storyturbtrump.chronopathy},
    we then use RTD at time scales of a day up to a year to
    quantify story turbulence and collective chronopathy.
    }
  \label{fig:storyturbtrump.allotaxonometer9000-trump-turbulence-year100-example}
\end{figure*}

\end{document}